# A COVID-19 Virus Epidemiological Model for Community and Policy Maker Use

Alex De Visscher

Department of Chemical and Materials Engineering, Gina Cody School of Engineering and Computer Science, Concordia University, Montreal, Quebec, Canada. Tel.: +1-514-848-2424 ext. 3488.

E-mail: alex.devisscher@concordia.ca

**Abstract**

An epidemiological model for COVID-19 was developed and implemented in MATLAB/GNU Octave for use by public health practitioners, policy makers and the general public. The model distinguishes four stages in the disease: infected, sick, seriously sick, and better. The model was preliminarily parameterized based on observations of the spread of the disease. The model is consistent with a mortality rate of 1.5 %. Preliminary simulations with the model indicate that concepts such as "herd immunity" and "flattening the curve" are highly misleading in the context of this virus. Public policies based on these concepts are inadequate to protect the population. Only reducing the $R_0$ of the virus below 1 is an effective strategy for maintaining the death burden of COVID-19 within the normal range of seasonal flu. As $R_0$ values estimated with the model range from 2.82 worldwide outside of China and 3.83 in the Western world in late February – early March 2020, this means social distancing with effectiveness greater than 65 % (worldwide) or 75 % (Western world) are needed to combat the virus successfully.

**Introduction**

The coronavirus disease 2019 (COVID-19) is a disease caused by SARS-CoV-2 (Severe Acute Respiratory Syndrome coronavirus 2). The disease broke out in Wuhan, Hubei, China, in December 2019 and caused a pandemic in the following months. Thanks to stringent control measures, the epidemic is nearly under control at the time of writing (March 2020) with approximately 81,000 cases and 3,200 deaths in China, and nearly 220,000 cases and 9,000 deaths worldwide as of March 18, 2020. Except for China and a number of southeast Asian countries, the pandemic is increasing exponentially in severity, with a doubling time of about 4 days in late February and the first half of March 2020. Doubling rates are substantially shorter in Europe and North America, generally less than 3 days.

Draconian measures were taken to control the spread of the disease in a number of Asian countries, but the rest of the world has been slow to follow suit despite the obvious dangers of delaying decisive action. I speculate that this is due in part to the lack of understanding of the mathematics of infectious diseases among policy makers and public health experts. Likewise, the public at large underestimates the stakes involved due to a lack of understanding of the impact of their own behavior.

The purpose of this paper is to present an epidemiological model that can be used by non-experts to explore the mathematics of the COVID-19 pandemic, and to present some preliminary results with the model. In the interest of time, no effort has been made to make the model fully accurate, or to optimize the parameterization of the model. Open sources such as the Worldometer Coronavirus website (https://www.worldometers.info/coronavirus) will be used as information source, to enable speedy development and publication. The model was developed in MATLAB and can be run with the open-source variant GNU Octave (https://www.gnu.org/software/octave).

**Model Development**

The mechanism assumed for the infection and spread of COVID-19 is shown in Figure 1. In the figure, U is the number of uninfected people, I is the number of infected people in the incubation period, S is the number of sick people, SS is the number of seriously sick people, D is the number of deceased, B is the number of people who are recovering, but not yet recovered ("better"), and R is the number of people who have completely recovered, and who are immune. The rates of transition from one state to another are indicated by $r_1$, $r_2$, etc. The rates are expressed in people per day.

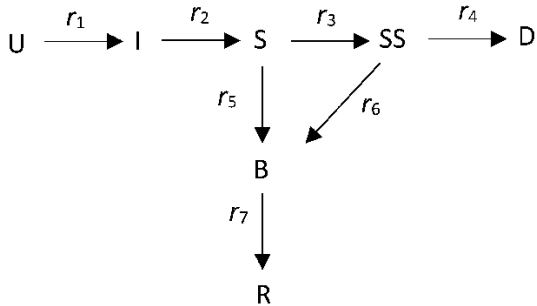

Figure 1. Mechanism of the COVID-19 model. U = uninfected, I = infected, S = sick, SS = seriously sick, D = dead, B = better, R = recovered; $r_1$ etc are rates of transition from one state to another (people per day)

The rate $r_1$, expressing the number of healthy, nonimmune people that are infected per day, is calculated based on the following assumptions:

- People in the categories I, S, SS and B can infect healthy people, each with a different rate.
- The infection rate is proportional to the fraction of people that are uninfected.

Based on these assumptions, the infection rate $r_1$ is calculated as follows:

$$r_1 = (k_{11}I + k_{12}S + k_{13}SS + k_{14}B)\frac{U}{P} \quad (1)$$

where $I$ is the number of people in category I, etc. $P$ is the number of people comprising the total population. $k_{11}$, $k_{12}$, $k_{13}$, and $k_{14}$ are rate constants (day$^{-1}$).

Next, it is assumed that all other transitions are first-order processes with rate constants $k_2$, $k_3$, etc., with rate constants expressed in day$^{-1}$. Hence:

$$r_2 = k_2 I \quad (2)$$

$$r_3 = k_3 S \quad (3)$$

$$r_4 = k_4 SS \quad (4)$$

$$r_5 = k_5 S \quad (5)$$

$$r_6 = k_6 SS \quad (6)$$

$$r_7 = k_7 B \quad (7)$$

Applying these rates to the mechanism in Figure 1, the dynamics of the COVID-19 pandemic can be modeled with the following differential equations:

$$\frac{dU}{dt} = -r_1 \quad (8)$$

$$\frac{dI}{dt} = r_1 - r_2 \quad (9)$$

$$\frac{dS}{dt} = r_2 - r_3 - r_5 \quad (10)$$

$$\frac{dSS}{dt} = r_3 - r_4 - r_6 \quad (11)$$

$$\frac{dD}{dt} = r_4 \quad (12)$$

$$\frac{dB}{dt} = r_5 + r_6 - r_7 \quad (13)$$

$$\frac{dR}{dt} = r_7 \quad (14)$$

where $t$ is the time in days.

**Parameterization**

To reduce the number of adjustable parameters, the following assumptions are made:

$$k_{12} = \frac{k_{11}}{2} \quad (15)$$

$$k_{13} = \frac{k_{11}}{3} \quad (16)$$

$$k_{14} = \frac{k_{11}}{4} \quad (17)$$

The remaining parameters are shown in Table 1.

Table 1. Parameters of the COVID-19 model

| parameter (units) | Value |
|---|---|
| $k_{11}$ (day$^{-1}$) | variable |
| $k_2$ (day$^{-1}$) | ln 2/5.1 |
| $k_3$ (day$^{-1}$) | $k_5$/9 |
| $k_4$ (day$^{-1}$) | $k_6 \times 15/85$ |
| $k_5$ (day$^{-1}$) | ln 2/3.5 |
| $k_6$ (day$^{-1}$) | ln 2/10 |
| $k_7$ (day$^{-1}$) | ln 2/3.5 |

$k_{11}$ was determined by trial and error, by comparing the doubling rate of the disease by observed values.

The value of $k_2$ is based on the observation that the median incubation time of COVID-19 is 5.1 days (Lauer et al., 2020).

The ratio $k_5/k_3$ (9) is based on the assumption that 10 % of the infected become seriously sick, whereas 90 % get better without developing serious symptoms. This is less than the observed proportion of roughly 80/20. The reason for the lower proportion assumed here is because many infected with mild symptoms remain undiagnosed, leading to an underreporting of mild cases.

The ratio $k_6/k_4$ is based on the assumption that 15 % of hospitalized COVID-19 patient do not survive.

The values of $k_5$ and $k_7$ are based on the assumption that the median duration of the disease is 3.5 days in the "sick" stage, followed by 3.5 days in the "better" stage. In other words, it is assumed that people developing mild symptoms recover in a week as a median value.

The value of $k_6$ is based on the assumption that the median seriously sick patient remains in this state for 10 days. Because there is a second pathway (dying), the actual median is less, 8.5 days. This is consistent with a mean duration of 12.26 days. This is in general agreement with clinical observations.

**Implementation**

The model was implemented in Matlab. The differential equations were integrated numerically with the 4-5$^{th}$ order Runge-Kutta-Fehlberg algorithm (function ode45 in Matlab). The following data and initial conditions were used:

$P$ = 100 million

$I$ = 100

$S$ = 10

$SS$ = 1

$D$ = 0

$B$ = 0

$R$ = 0

At time zero there are 111 infected people among a population of 100 million. In this early phase it can be assumed that the number of known infections will be on the order of 10 or less. In other words, we are starting the simulation very early on. The doubling time is calculated from the total number of people in all infected stages on day 29 and day 30 with the equation:

$$t_{\text{double}} = \frac{\ln(2)}{\ln\left(\frac{C_{30}}{C_{29}}\right)} \qquad (18)$$

where $C_n$ is an estimate of the number of known or reported "cases" on day $n$. The number of known cases is assumed to be 5 % of infected, a third of sick, 90 % of seriously sick, 12 % of recovering, 12 % of recovered, and 90 % of deceased patients. The calculated doubling time is not sensitive to the choice of these fractions.

To calculate the value of $R_0$ (the number of people infected by the average carrier of the virus), a separate simulation was run with a single infected patient, where the number of newly infected is calculated over time.

**Results**

*Doubling Times, Infection Rates, Infection Factors*

First, the doubling time of the pandemic is calculated for different values of the infection rate $k_{11}$. The result is shown in Figure 2. The figure is limited to parameter values that lead to $R_0$ values greater than 1, where the doubling time can be expected to be approximately constant in time.

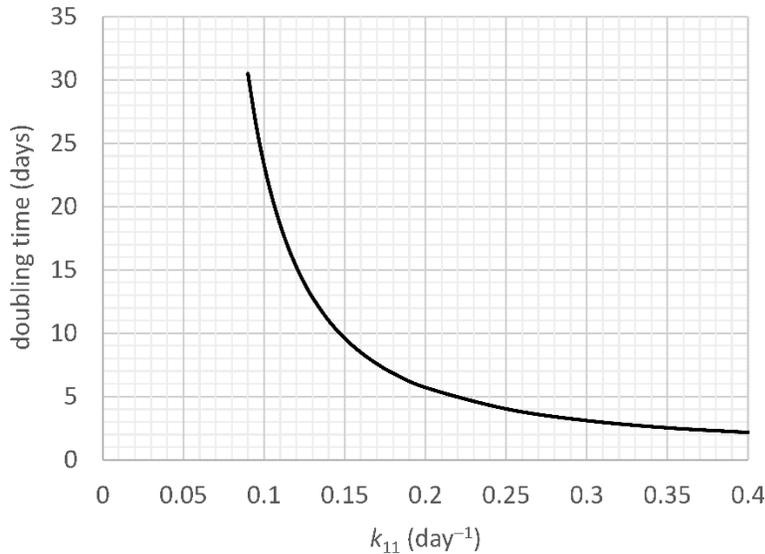

Figure 2. Doubling time versus infection rate ($k_{11}$)

As mentioned in the Introduction, the worldwide doubling time of COVID-19 outside China was 4 days in the latter half of February and the first half of March 2020. This corresponds with $k_{11} = 0.25$ day$^{-1}$. This value was used as the default in further simulations, unless specified otherwise.

In Europe and North America, doubling times were significantly shorter during that time. For instance, in Italy, the reported number of COVID-19 cases grew approximately exponentially from 150 on February 23 to 10,149 on March 10 (16 days later). An exponential fit to the data leads to a doubling time of 2.66 days ($R^2 = 0.9841$). This is consistent with $k_{11} = 0.34$ day$^{-1}$. Most of the Western world experienced similar growth rates during the same time.

The $R_0$ value is calculated as a function of $k_{11}$ as well. The results are show in Figure 3. The $R_0$ value reaches 1 when $k_{11} = 0.089$ day$^{-1}$, only 35.6 % of the global average $k_{11}$ value in late February to early March 2020, and 26.1 % of the value in Italy during that time. As a result, the model-based estimate for $R_0$ in late February to early March is 2.82 worldwide outside China, and 3.83 in Italy and most of the Western world.

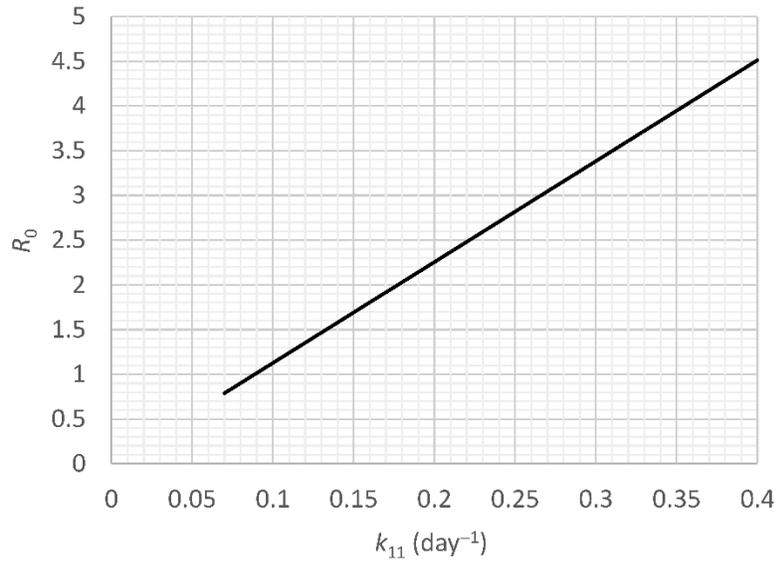

Figure 3. $R_0$ value versus infection rate ($k_{11}$)

The relationship between $R_0$ and $k_{11}$ follows a perfect linear relationship as follows:

$$R_0 = k_{11} \times 11.28 \text{ days} \qquad (19)$$

The doubling time is the variable that is generally accessible rather than the infection rate, so for convenience, a chart of the $R_0$ value versus doubling time was generated as well, to enable fast evaluation without the need to run the model. This is shown in Figure 4. The relationship is only meaningful for $R_0$ values significantly in excess of 1.

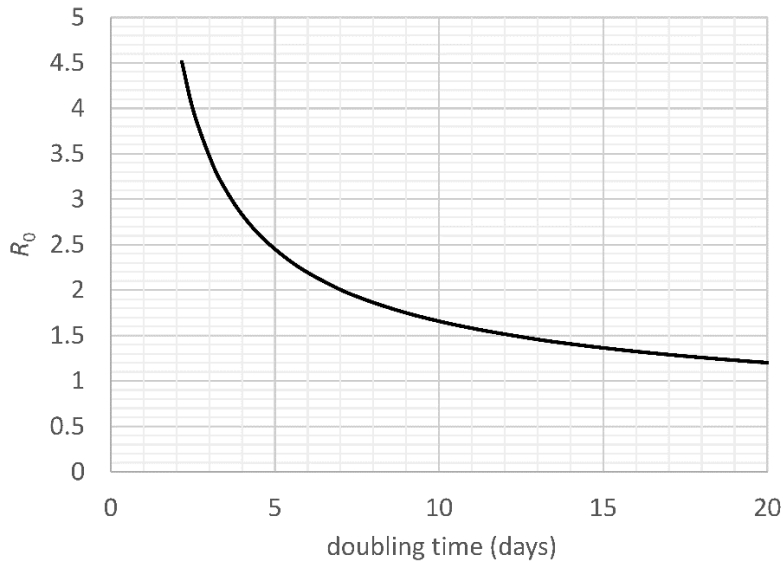

Figure 4. $R_0$ value versus doubling time

*Scenarios – Average, Fast, Slow*

In this section, a number of scenarios will be run to assess the number of infected and the number of deaths as a function of time, for a population of 100 million, starting with 111 infected (100 incubating, 10 sick and 1 seriously sick) at time 0.

Figures 5 and 6 show the evolution of the epidemic in the base case (doubling time = 4 days, $k_{11}$ = 0.25 day$^{-1}$, $R_0$ = 2.82), without intervention. The first deaths are predicted around day 12, when about 1000 people are infected. The number of people showing symptoms at this time is around 390 (240 mild, 20 serious, 130 recovering). This early in the epidemic, it is likely that testing is not yet fully deployed, and the number of reported cases is likely to be on the order of 200 or less.

After one month, the model predicts about 30 deaths and a total of about 24,000 infected. Of these, about 15,000 show no symptoms, 5,000 show mild symptoms, 500 severe symptoms, and 3000 are recovering. At this point, the official case count is probably a few thousands. Around this time, most governments start taking serious precautions to limit the spread of the virus.

After two months without intervention, there are 3.9 million infections and almost 5,600 deaths. As a rule of thumb, there is one death per 700 cases in the expansion phase of the disease when the doubling time is 4 days. 2.46 million people are in the incubation phase and 77,000 people are seriously sick.

The peak of seriously sick people is reached on day 96, when over 2.6 million people are seriously sick and over half a million people have died.

After 150 days, the disease is declining but is still overwhelming the health care system, with almost 200,000 people seriously sick. The model predicts 1.35 million deaths at this time, 1.35 % of the population. Given the severe lack of care that would occur, the death toll is probably underestimated by a factor 2 or 3. About 92.6 million people get infected overall, significantly more than the expected number from "herd immunity" (64.4 million). This is because the disease expands so rapidly that it overshoots and continues to infect people as it winds down past the 65 million mark. This simulation clearly shows that herd immunity is only effective when people are vaccinated before the spread of the disease.

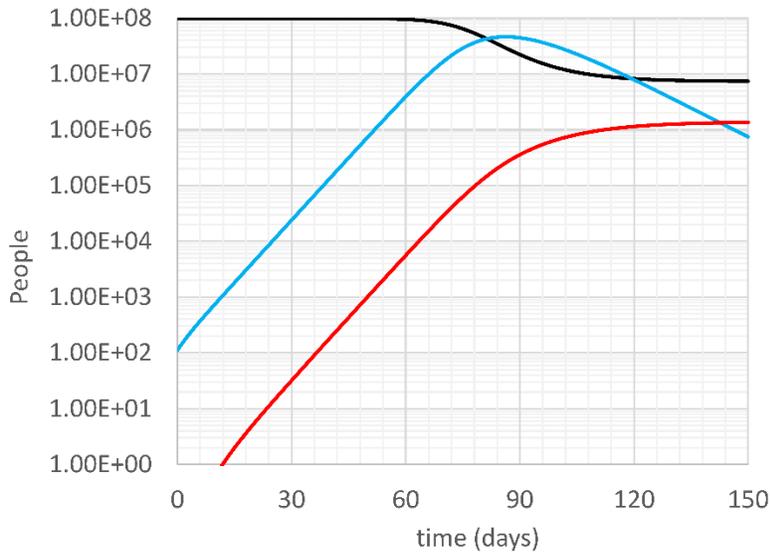

Figure 5. Uninfected (black), all infected (blue) and deceased (red) people versus time. Base case, no intervention, doubling time 4 days

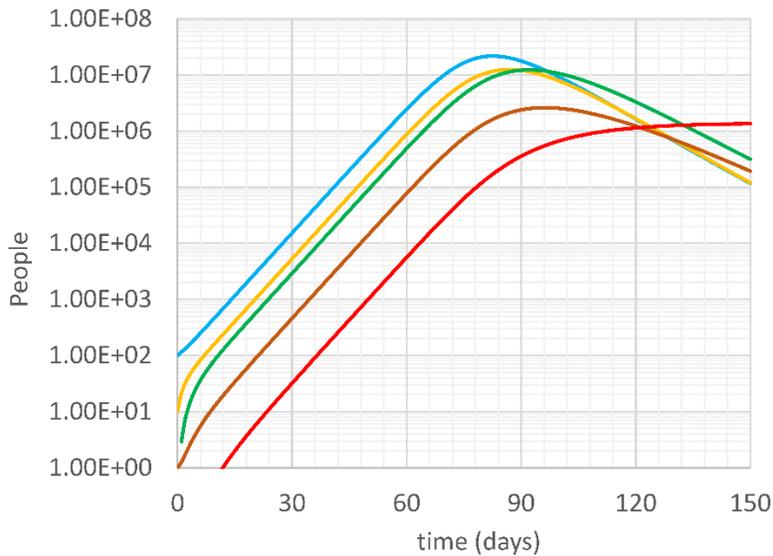

Figure 6. Incubating (blue), sick (yellow), seriously sick (brown), recovering (green) and deceased (red) people versus time. Base case, no intervention, doubling time 4 days

arXiv:2003.08824　　　　　　　　　　　　　　　　　　　　　　　　　　　　　　　　　　　　　　　　　　9

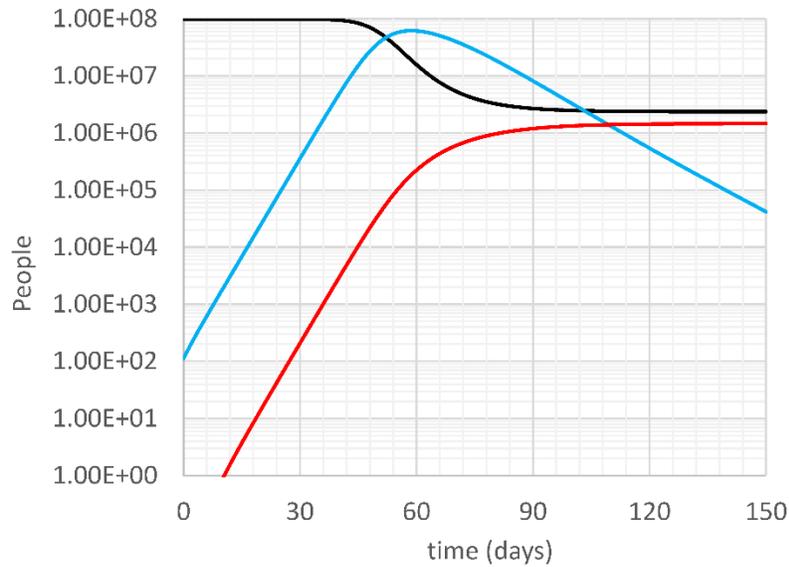

Figure 7. Uninfected (black), all infected (blue) and deceased (red) people versus time. Fast case, no intervention, doubling time 2.66 days

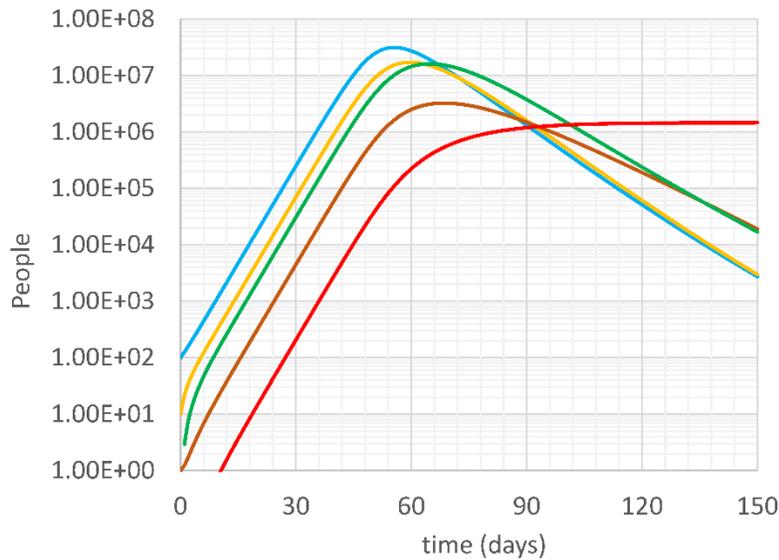

Figure 8. Incubating (blue), sick (yellow), seriously sick (brown), recovering (green) and deceased (red) people versus time. Fast case, no intervention, doubling time 2.66 days

Next, the simulation was repeated for a "fast" scenario where the doubling time is the same as in Italy in late February to early March, 2.66 days ($k_{11}$ = 0.34 day$^{-1}$, $R_0$ = 3.83). The results are shown in Figures 7 and 8. The main difference with the base case is that the disease spreads faster and peaks sooner. At its peak, 3.25 million people are seriously sick, on day 69. The death burden after 150 days is 1.46 million, or 1.46

% of the population. At this time, the disease has affected 97.6 million people, 97.6 % of the population. Again, this is massively above the number expected from herd immunity (73.9 million people).

During the initial spread of the disease, there is one death per 1700 cases, indicating that the epidemic may be underestimated even more when it spreads rapidly.

The next scenario represents a strategy that is popularized as "flattening the curve": the infection rate is significantly reduced to slow down the spread of the disease in an attempt to avoid overburdening the health care system, but no attempt is made to eradicate the disease, i.e., the $R_0$ remains significantly above 1. The simulation is run with an infection rate $k_{11} = 0.16$ day$^{-1}$ (doubling time 8.48 days, $R_0 = 1.81$). The result is shown in Figures 9 and 10.

The peak in the number of seriously sick people is significantly delayed, to day 185, but the number of patients still far exceeds the capacity of any health care system, with 1.3 million seriously sick, half the number of the base case. The death burden in the "flattening the curve" strategy is slightly over 1 million, still over two-thirds of the fast scenario. The total number of people that get infected in a 240-day time span is 72.4 million, again markedly more than the number expected from herd immunity considerations (44.8 million).

Clearly, flattening the curve is an inadequate strategy for fighting the COVID-19 pandemic.

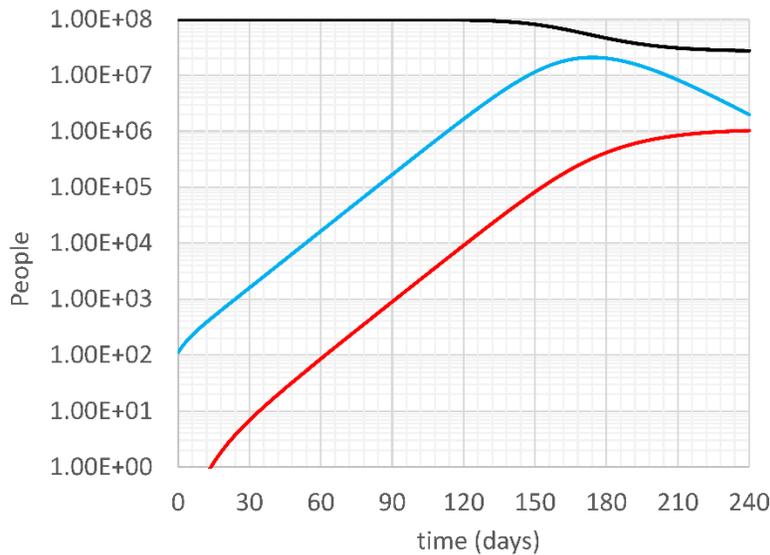

Figure 9. Uninfected (black), all infected (blue) and deceased (red) people versus time. Slow, "flattening the curve" case, no intervention, doubling time 8.48 days

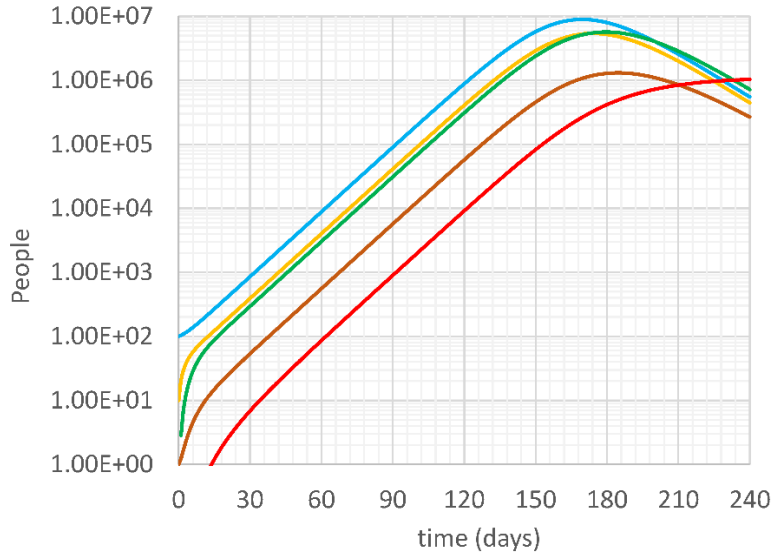

Figure 10. Incubating (blue), sick (yellow), seriously sick (brown), recovering (green) and deceased (red) people versus time. Slow, "flattening the curve" case, no intervention, doubling time 8.48 days

*Scenarios – Social Distancing Intervention*

Next, starting from the base case, it is assumed that drastic social distancing measures are taken on day 30 that reduce $R_0$ to below 1. It is assumed that the value of $k_{11}$ is reduced by 70 % (i.e., from 0.25 day$^{-1}$ to 0.075 day$^{-1}$ i.e., $R_0$ decreases from 2.82 to 0.85). The change of $k_{11}$ is introduced gradually over a couple of days, centered around day 30, to avoid numerical difficulties in the computer simulation. The result is shown in Figures 11 and 12. A 70 % effective social distancing intervention with a starting value of $k_{11}$ = 0.25 day$^{-1}$, i.e., with respect to the world average, is equivalent with a 78 % effective intervention with a starting value of $k_{11}$ = 0.34 day$^{-1}$, i.e., with respect to the situation in Italy and most of the Western world. In other words, in much of the Western world, the results shown below reflect a social distancing initiative that is 78 % effective, not 70 %.

There is a marked decline in the number of infected in this scenario. After 240 days, the number of people who died of COVID-19 is 1770, almost three orders of magnitude less than the previous scenarios. Still, this number is 55 times the number people who had died at the onset of the intervention (32). An additional 50 people pass away between day 240 and day 300.

The number of seriously sick people peaks at a value of 1589 on day 53, again about thee orders of magnitude less than in the preceding scenarios.

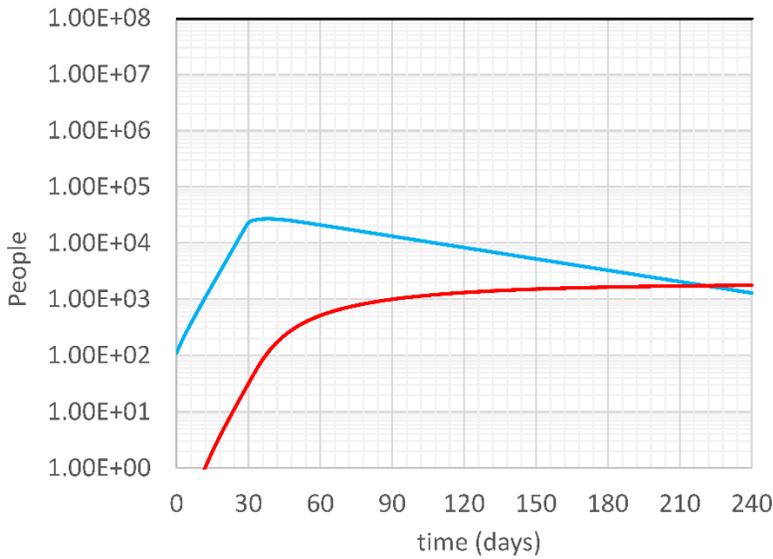

Figure 11. Uninfected (black), all infected (blue) and deceased (red) people versus time. Base case, doubling time 4 days, intervention on day 30 with 70 % effectiveness

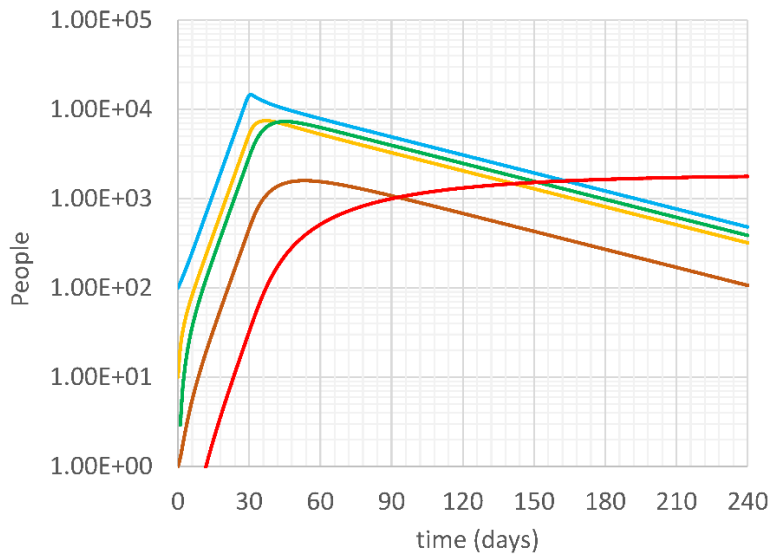

Figure 12. Incubating (blue), sick (yellow), seriously sick (brown), recovering (green) and deceased (red) people versus time. Base case, doubling time 4 days, intervention on day 30 with 70 % effectiveness

What is clear from this scenario is that the decline of the epidemic is much slower than its growth. This has important repercussions for any public health policy aiming to save lives. Even seven months into the intervention, the number of infected is comparable to the number of infected two and a half weeks before the intervention. Terminating the intervention would immediately relaunch the epidemic. The intervention must be maintained until the population can be vaccinated on a large scale.

arXiv:2003.08824　　　　　　　　　　　　　　　　　　　　　　　　　　　　　　　　　　　　　　　　　　13

*Scenarios – The Death Burden of Inaction*

In this section, the number of deaths will be evaluated as a function of time and effectiveness of the social distancing intervention. The starting point is the base case, with a doubling time of 4 days ($k_{11}$ = 0.25 day$^{-1}$, $R_0$ = 2.82).

First the effect of effectiveness of the social distancing intervention is calculated. It is assumed that the intervention starts on day 30 with an effectiveness ranging from 50 % to 80 %. Figure 13 shows the number of deaths after 60, 150, and 300 days.

After 60 days, i.e., 30 days after the start of the intervention, the effect of effectiveness of intervention on mortality is very limited. This is concerning because to observers it appears that the interventions are not working. However, over a 150-day time span, a 5 % decrease of efficiency can triple the mortality. Over a 300-day time span, a 1 % decrease of the efficiency (e.g., from 62 % to 61 %) can cause a 50 % increase in mortality. This explains why some Asian countries treat seemingly trivial violations of the social distancing rules as felonies.

The value of $R_0$ equals 1 at 64.4 % efficiency in this case. The importance of keeping $R_0$ below 1 is immediately obvious from Figure 13. When the initial value of $k_{11}$ is 0.34 day$^{-1}$, an efficiency of 73.9 % is needed to lower $R_0$ to 1. This should be the minimum target efficiency of social distancing in Europe and North America.

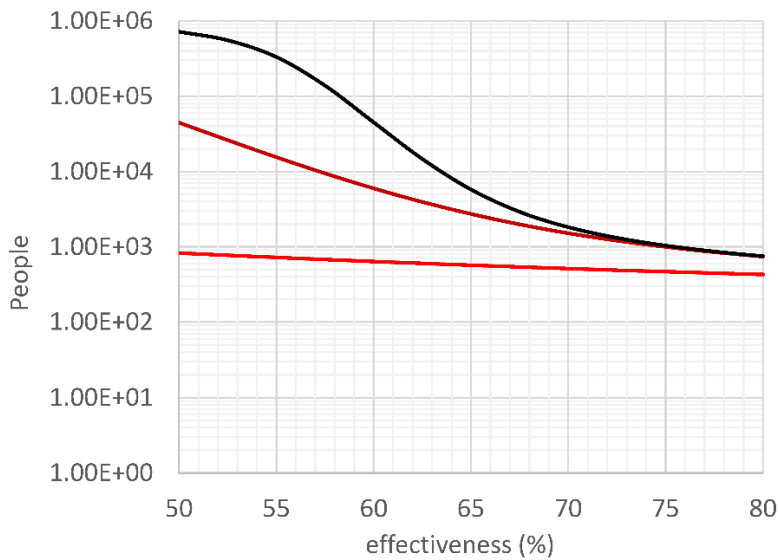

Figure 13. Number of deaths after 60 days (bright red), 150 days (dark red) and 300 days (black) versus effectiveness of intervention. Base case, doubling time 4 days, intervention on day 30

Next, the effect of timing of introduction of a social distancing intervention on the mortality over 60 days, 150 days, and 300 days is calculated. The results are shown in Figure 14. Probably not surprisingly, the number of deaths doubles with every 4-day delay of the introduction of social distancing. This is an important point, because the number of deaths may seem small at the time of introduction (e.g., from 32 on day 30 to 64 on day 34), the number of deaths after 300 days increases from 1,822 to 3,607 as a result of this delay. Every additional death at the time of intervention represents 55 additional deaths over a 300-day time span.

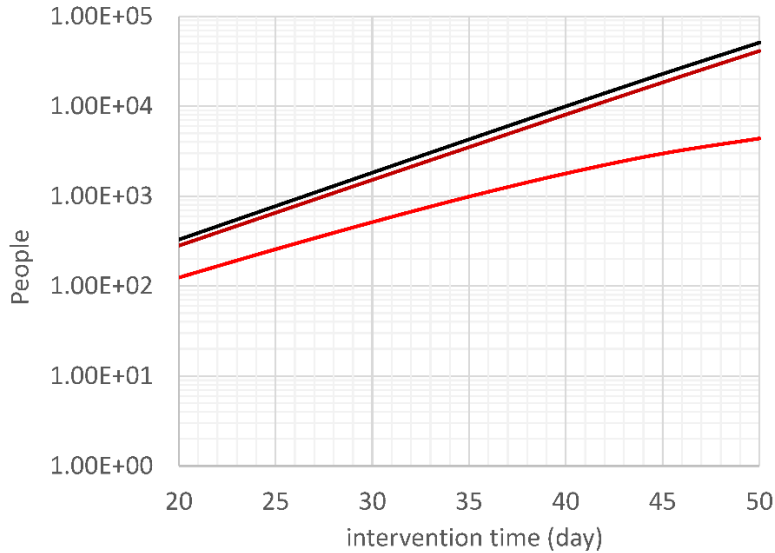

Figure 14. Number of deaths after 60 days (bright red), 150 days (dark red) and 300 days (black) versus time of intervention. Base case, doubling time 4 days, intervention effectiveness 70 %

**Discussion**

The worldwide average $R_0$ value of COVID-19 outside China is estimated at 2.82 for the late February – early March 2020 period, based on a doubling time of 4 days, whereas the value was around 3.83 at the same time in the Western world, based on a doubling time of 2.66 days. This is significantly higher than regular flu viruses. As a result, experience with flu is a poor guide for predicting the course of the COVID-19 epidemic. For instance, flu viruses are seasonal because their $R_0$ tend to drop below 1 over the summer months, but COVID-19 is too contagious to display a similar seasonality.

A doubling time of 4 days is consistent with an infection rate $k_{11}$ of 0.25 day$^{-1}$. This value is specific to the current model and should not be applied to other models. It should also not be confused with the apparent growth rate of the virus.

Based on the chosen parameter values, the disease intrinsically has a mortality of 1.5 %, which is within the range of estimates that are going around at the time of writing. The WHO proposes a mortality of 3.4 % whereas the mortality in South Korea is 1 % and rising at the time of writing. On the other hand, the mortality in Italy is almost 10 %, probably as a result of underreporting of mild cases and overreporting of deaths attributed to COVID-19.

The model predicts that there is 1 death per 700 cases during the growth phase of the epidemic when the doubling time is 4 days, and 1 death per 1700 cases when the doubling time is 2.66 days. With nearly 6000 deaths outside of China as of March 18, this means that the number of infected outside of China is probably on the order of 4 million people, about 30 times the number of reported cases. Assuming a death rate of 1.5 %, a lower limit of 60,000 people outside of China can be expected even if no new infections occur.

Irrespective of the variables used, model predictions indicate that COVID-19 will affect vastly more people than expected from "herd immunity" considerations. This is because there is a huge number of infected people at the time of onset of herd immunity, enough to infect most of the remaining uninfected people before the epidemic spirals down. It follows that public health strategies based on herd immunity are extremely misguided and extremely deadly. Herd immunity is only effective when the population is vaccinated before the onset of the disease.

Likewise, a public health strategy based on "flattening the curve" without diminishing $R_0$ below 1 is inadequate and extremely deadly, with an expected mortality rate of about 1 % based on the entire population even with the unrealistic assumption that the healthcare system can handle the number of patients.

Once a country takes decisive action to reduce $R_0$ below 1, the mortality still increases by about a factor 55 before the disease is stopped. Based on that number, and assuming that the world's governments take appropriate action at the time of writing, the mortality outside China can be expected to be 330,000. Some governments have already taken decisive action but many have not, so it is very likely that the actual death burden of COVID-19 will be significantly higher. Every four days of delay can be expected to double that number.

The mortality of COVID-19 after intervention is very sensitive to the effectiveness of the intervention. For instance, lowering the effectiveness from 65 % to 60 % increases the mortality by a factor 8, from about 55 per million to about 450 per million in the case investigated. Minor gaps in the social distancing policy (e.g., closing bars but allowing private parties), or a small fraction of the population violating the policy can have disastrous effects on mortality rates. Lax enforcement of a social distancing policy could be costly in lives.

**Software**

The software of the model consists of two MATLAB files, main.m and f.m. The file main.m is the control file that should be run. The file f.m defines the differential equations. The model can be run on MATLAB, or on its open-source equivalent GNU Octave (https://www.gnu.org/software/octave). The source code is shown in Appendix, and can be obtained from the author by e-mail.

Two additional files are included for the calculation of $R_0$: main_R0.m and fR0.m.

For terms of use: see source code.

**Conclusions**

Calculations with an epidemiological model developed to describe the spread of the COVID-19 pandemic indicates that the mortality of the pandemic will be over 1 % of the global population even if moderate social distances "flattening the curve" are imposed. This does not account for overburdening of the healthcare system and associated deaths. Likewise, modeling shows that it is a fatal misconception to assume that "herd immunity" can provide even the slightest protection against the virus in the current circumstances.

The mortality of COVID-19 can be reduced by up to three orders of magnitude, at least in principle, by reducing $R_0$ below 1 early on in the development of the pandemic. However, failure to meet that objective has a huge impact on mortality, up to 50 % increase for a 1 % decrease in social distancing effectiveness. Every four days of delay in decisive action can double the mortality in the long run.

**Appendix: Source Codes**

**`main.m`**

```matlab
% COVID-19 epidemiological model for public, health practitioners and
% policymakers
% Written by Alex De Visscher in March 2020
% -----------------------------------------
% Users have permission to use the model for noncommercial purposes
% Users have permission to publish results obtained with the model,
% provided that they cite the ArXiv paper associated with the model
% Any changes to the model must be reported to the author at
% alex.devisscher@concordia.ca
% The author does not permit any changes to the model that have the purpose
% of generating inaccurate or misleading results.
% While all efforts were made to ensure that the model is valid and
% accurate, the author cannot be held responsible for any damage, loss, or
% injury caused by the use of this model. This model does not replace
% expert opinion on any public health matter.
% -----------------------------------------
clear all
tspan = 0:300   % time span of simulation (days)
% Rate constants (1/day):
```

```matlab
k11 = 0.25;
k12 = k11/2;
k13 = k11/3;
k14 = k11/4;
k2 = log(2)/5.1;
k5 = log(2)/3.5;
k6 = log(2)/10;
k7 = log(2)/3.5;
k3 = k5/9;
k4 = k6*15/85;
P = 1e8;        % total population
% Initial conditions:
I = 100;        % infected, in the incubation phase
S = 10;         % sick
SS = 1;         % seriously sick
B = 0;          % better
R = 0;          % recovered
D = 0;          % dead
U = P - I - S - SS;    % uninfected
y0(1) = U;
y0(2) = I;
y0(3) = S;
y0(4) = SS;
y0(5) = B;
y0(6) = R;
y0(7) = D;
interv_time = 30;    % intervention time; > maximum time if no intervention
interv_success = 0.7; % fractional reduction of k11 during intervention
options = odeset('RelTol', 1e-6, 'AbsTol', 1e-8, 'InitialStep', 0.01);
[T,Y] =
ode45(@f,tspan,y0,options,k11,k12,k13,k14,k2,k3,k4,k5,k6,k7,P,interv_time,interv_success);
[s1,s2] = size(T);
figure(1)
plot(T,Y(1:s1,2:7))
T
Y
jcomp=4; %Component number
figure(2)
semilogy(T,Y(1:s1,jcomp))
Y(1:s1,jcomp)
for j = 1:s1
    % 1 = all cases; 2 = reported cases; 3 = serious cases reported;
    % 4 = deaths reported; 5 = actual deaths
    cases(j,1) = Y(j,2)+Y(j,3)+Y(j,4)+Y(j,5)+Y(j,6)+Y(j,7);
    cases(j,2) =
0.05*Y(j,2)+0.3333*Y(j,3)+0.9*Y(j,4)+0.12*Y(j,5)+0.12*Y(j,6)+0.9*Y(j,7);
    cases(j,3) = 0.9*Y(j,4);
    cases(j,4) = 0.9*Y(j,7);
    cases(j,5) = Y(j,7);
end
figure(3)
semilogy(T(2:s1),cases(2:s1,1:5))
frac = cases(s1,2)/cases(s1-1,2);
k_growth = log(frac);
doubling_time = log(2)/k_growth
final_deaths = Y(s1,7)
total_infections = Y(s1,2)+Y(s1,3)+Y(s1,4)+Y(s1,5)+Y(s1,6)+Y(s1,7)
```

**f.m**

```matlab
% Function calculating the differential equations for the COVID-19 model
% Written by Alex De Visscher in March 2020
% For terms of use: see main file (main.m)
% ----------------------------------------
function dydt = f(t,y,k110,k120,k130,k140,k2,k3,k4,k5,k6,k7,P,interv_time,interv_success)

dydt = zeros(7,1);
U = y(1);
I = y(2);
S = y(3);
SS = y(4);
B = y(5);
R = y(6);
D = y(7);
interv_correction = 1 - interv_success/2 - (interv_success/2)*erf(t-interv_time);
k11 = k110*interv_correction;
k12 = k120*interv_correction;
k13 = k130*interv_correction;
k14 = k140*interv_correction;
r1 = k11*I + k12*S + k13*SS + k14*B;
r1 = r1*U/P;
r2 = k2*I;
r3 = k3*S;
r4 = k4*SS;
r5 = k5*S;
r6 = k6*SS;
r7 = k7*B;
dUdt = -r1;
dIdt = r1 - r2;
dSdt = r2 - r3 - r5;
dSSdt = r3 - r4 - r6;
dDdt = r4;
dBdt = r5 + r6 - r7;
dRdt = r7;
dydt(1) = dUdt;
dydt(2) = dIdt;
dydt(3) = dSdt;
dydt(4) = dSSdt;
dydt(5) = dBdt;
dydt(6) = dRdt;
dydt(7) = dDdt;
```

**main_R0.m**

```matlab
% COVID-19 epidemiological model for public, health practitioners and
% policymakers - calculation of R0
% Written by Alex De Visscher in March 2020
% Terms of use: see file main.m
% -----------------------------
clear all
tspan = 0:150
k11 = 0.25;
```

arXiv:2003.08824      19

```matlab
k12 = k11/2;
k13 = k11/3;
k14 = k11/4;
k2 = log(2)/5.1;
k5 = log(2)/3.5;
k6 = log(2)/10;
k7 = log(2)/3.5;
k3 = k5/9;
k4 = k6*15/85;
P = 1e8;
I = 1;
U = P - I;
S = 0;
SS = 0;
B = 0;
R = 0;
D = 0;
y0(1) = U;
y0(2) = I;
y0(3) = S;
y0(4) = SS;
y0(5) = B;
y0(6) = R;
y0(7) = D;
y0(8) = 0;
interv_time = 320;
interv_success = 0.7;
options = odeset('RelTol', 1e-6, 'AbsTol', 1e-8, 'InitialStep', 0.01);
[T,Y] =
ode45(@fR0,tspan,y0,options,k11,k12,k13,k14,k2,k3,k4,k5,k6,k7,P,interv_time,interv_suc
cess);
[s1,s2] = size(T);
figure(1)
plot(T,Y(1:s1,2:7))
T
Y
jcomp=4; %Component number
figure(2)
semilogy(T,Y(1:s1,jcomp))
Y(1:s1,jcomp)
for j = 1:s1
    % 1 = all cases; 2 = reported cases; 3 = serious cases reported;
    % 4 = deaths reported; 5 = actual deaths
    cases(j,1) = Y(j,2)+Y(j,3)+Y(j,4)+Y(j,5)+Y(j,6)+Y(j,7);
    cases(j,2) =
0.05*Y(j,2)+0.3333*Y(j,3)+0.9*Y(j,4)+0.12*Y(j,5)+0.12*Y(j,6)+0.9*Y(j,7);
    cases(j,3) = 0.9*Y(j,4);
    cases(j,4) = 0.9*Y(j,7);
    cases(j,5) = Y(j,7);
end
figure(3)
semilogy(T(2:s1),cases(2:s1,1:5))
%frac = cases(s1,2)/cases(s1-1,2);
%k_growth = log(frac);
%doubling_time = log(2)/k_growth
death_rate = Y(s1,7)
R0 = Y(s1,8)
```

## fR0.m

```matlab
% COVID-19 epidemiological model for public, health practitioners and
% policymakers - calculation of R0
% Written by Alex De Visscher in March 2020
% Terms of use: see file main.m
% -----------------------------
function dydt = f(t,y,k110,k120,k130,k140,k2,k3,k4,k5,k6,k7,P,interv_time,interv_success)

dydt = zeros(8,1);
U = y(1);
I = y(2);
S = y(3);
SS = y(4);
B = y(5);
R = y(6);
D = y(7);
interv_correction = 1 - interv_success/2 - (interv_success/2)*erf(t-interv_time);
k11 = k110*interv_correction;
k12 = k120*interv_correction;
k13 = k130*interv_correction;
k14 = k140*interv_correction;
r1 = k11*I + k12*S + k13*SS + k14*B;
r1 = r1*U/P;
r2 = k2*I;
r3 = k3*S;
r4 = k4*SS;
r5 = k5*S;
r6 = k6*SS;
r7 = k7*B;
dUdt = 0;
dIdt =  - r2;
dSdt = r2 - r3 - r5;
dSSdt = r3 - r4 - r6;
dDdt = r4;
dBdt = r5 + r6 - r7;
dRdt = r7;
dydt(1) = dUdt;
dydt(2) = dIdt;
dydt(3) = dSdt;
dydt(4) = dSSdt;
dydt(5) = dBdt;
dydt(6) = dRdt;
dydt(7) = dDdt;
dydt(8) = r1;
```